\documentclass{article}

\usepackage{arxiv}

\usepackage[utf8]{inputenc} % allow utf-8 input
\usepackage[T1]{fontenc}    % use 8-bit T1 fonts
\usepackage{hyperref}       % hyperlinks
\usepackage{url}            % simple URL typesetting
\usepackage{booktabs}       % professional-quality tables
\usepackage{amsfonts}       % blackboard math symbols
\usepackage{nicefrac}       % compact symbols for 1/2, etc.
\usepackage{microtype}      % microtypography
\usepackage{lipsum}		% Can be removed after putting your text content
\usepackage{graphicx}
\usepackage{natbib}
\usepackage{doi}

\usepackage{multirow}
\usepackage{amsmath}
\usepackage{amssymb}
\usepackage{latexsym}
\usepackage{placeins}

\newcommand{\mb}[1]{\ensuremath{\mathbf{#1}}}
\newcommand{\mc}[1]{\ensuremath{\mathcal{#1}}}
\newcommand{\mr}[1]{\ensuremath{\mathrm{#1}}}

\title{Benchmarking common uncertainty estimation methods with histopathological images under domain shift and label noise}

\date{September 2022} 					% Or removing it

\author{ Hendrik Mehrtens\footnotemark[1], Alexander Kurz\footnotemark[1], Tabea-Clara Bucher, Titus J. Brinker\footnotemark[2]\\
	Division of Digital Biomarkers for Oncology \\ German Cancer Research Center (DKFZ) \\ Heidelberg, Germany \\
	\texttt{\{hendrikalexander.mehrtens\},\{alexander.kurz\},\{tabea.bucher\}@dkfz-heidelberg.de} \\
	\texttt{titus.brinker@nct-heidelberg.de} \\
}

\renewcommand{\headeright}{Mehrtens et al., 2022}
\renewcommand{\undertitle}{Preprint}
\renewcommand{\shorttitle}{Benchmarking Uncertainty in Histopathology}

\hypersetup{
pdftitle={A template for the arxiv style},
pdfsubject={q-bio.NC, q-bio.QM},
pdfauthor={David S.~Hippocampus, Elias D.~Striatum},
pdfkeywords={First keyword, Second keyword, More},
}

\begin{document}
\maketitle

\footnotetext[1]{Contributed equally}
\footnotetext[2]{Corresponding author}

\begin{abstract}
In the past years, deep learning has seen an increase in usage in the domain of histopathological applications. However, while these approaches have shown great potential, in high-risk environments deep learning models need to be able to judge their uncertainty and be able to reject inputs when there is a significant chance of misclassification. 
In this work, we conduct a rigorous evaluation of the most commonly used uncertainty and robustness methods for the classification of Whole Slide Images, with a focus on the task of selective classification, where the model should reject the classification in situations in which it is uncertain. 
We conduct our experiments on tile-level under the aspects of domain shift and label noise, as well as on slide-level. In our experiments, we compare Deep Ensembles, Monte-Carlo Dropout, Stochastic Variational Inference, Test-Time Data Augmentation as well as ensembles of the latter approaches.
We observe that ensembles of methods generally lead to better uncertainty estimates as well as an increased robustness towards domain shifts and label noise, while contrary to results from classical computer vision benchmarks no systematic gain of the other methods can be shown. 
Across methods, a rejection of the most uncertain samples reliably leads to a significant increase in classification accuracy on both in-distribution as well as out-of-distribution data. Furthermore, we conduct experiments comparing these methods under varying conditions of label noise.
Lastly, we publish our code framework to facilitate further research on uncertainty estimation on histopathological data.

\end{abstract}

% keywords can be removed

\section{Introduction}
\label{sec1:introduction}

Deep Neural Networks (DNNs) have been shown to be of equal or even superior performance in studies on many medical tasks \citep{esteva2017, haenssle2018, hekler2019}, compared to human practitioners. Nonetheless, they have rarely been adopted in clinical practice. One commonly given reason \citep{begoli2019, kompa2021a, vanderlaak2021} is their inability to provide well-calibrated estimates of their predictive uncertainty  \citep{guo2017}, thereby prohibiting the practitioner from judging the reliability of the system's decision, which is a necessary condition in areas of high uncertainty like medical decision-making.

Ideally, a well-calibrated deep learning system should be able to judge and communicate a correct estimate of its certainty in each prediction, not only informing the human practitioner of its momentary reliability but also enabling the system to automatically reject inputs \citep{band2021, jaeger2023} and refer them to humans for inspection, thereby enhancing the reliability of the system.

Another problem in practice is the high vulnerability of DNN-based systems to "domain shifts", which are differences between the data distributions in the training and deployment setting. In the general machine learning literature, \citep{hendrycks2019, ovadia2019} noticed the vulnerability of deep neural networks to even slight artificial perturbations of images, which result in drastic deterioration of classification performance. In the area of digital pathology, these kinds of shifts are very common \citep{stacke2021}. They arise due to different data-acquisition processes across clinics, for example, due to different staining procedures or different scanners but can also be caused by changes in the distribution of patient characteristics (gender, age, etc.). The unpredictable behavior of models in new data regimes limits the ability to put confidence in a system's decision, especially in high-risk environments, an example being decisions that affect a patient's treatment.

Traditional methods for improving the generalization of a machine learning system use augmentations of the input data, which is currently heavily investigated in the field of digital histopathology \citep{tellez2019}.
In recent years, several new techniques for uncertainty quantification and generalization under domain shift have been proposed. Yet, most of these methods have primarily been evaluated in the general setting of computer vision and object detection, with limited research on them specifically in the field of digital pathology.

Whole Slide Images (WSIs) are a challenging domain for deep learning, due to the large size of the images, typically at the gigapixel scale, and the limited availability of labeled data. Additionally, as the evaluation of a WSI is error-prone, with high inter-observer variability \citep{karimi2020a}, labels can be subject to a high degree of label noise, making the integration of predictive certainty vital.

\citep{linmans2020} explored the capability of multi-headed ensemble networks to detect out-of-distribution (OOD) inputs in WSIs. \citep{thagaard2020} compared the ability of Deep Ensembles \citep{lakshminarayanan2017} and Monte-Carlo Dropout (MCDO) \citep{gal2016} to estimate predictive uncertainty at the tile-level. Independently and concurrent to our work, \citep{linmans2023} compared prominent methods for uncertainty estimation on histopathological slides, with a focus on out-of-distribution detection of foreign tissue in breast and prostate tissues. Building upon Jaeger's \citep{jaeger2023} insights on meaningful and comparable evaluation of uncertainty estimation methods, we investigate the ability of uncertainty estimation methods to reject uncertain predictions. Our evaluations are performed at the tile-level, as well as at the slide-level, which is the clinically more relevant task.

We carefully compare the robustness of Deep Ensembles \citep{lakshminarayanan2017}, MCDO \citep{gal2016}, Stochastic Variational Inference (SVI) \citep{blundell2015}, Test-Time Data Augmentation (TTA) \citep{ayhan2018} and ensembles of the latter approaches under domain shift. We study the impact of the reject option by uncertainty on the performance of the model, observing a significant increase in accuracy in high-certainty predictions, and compare multiple metrics of estimating the final certainty from the predictions. We simulate the effects of label noise in the edge regions of the tumor annotations and measure the robustness of the methods to this induced label noise.

For our experiments, we use multiple tumorous histopathological datasets. For our tile-level experiments, we utilize the Camelyon17 \citep{bandi2019} dataset. The Camelyon17 breast cancer tissue dataset comprises a total of 50 slides with annotated tumor regions. These slides were obtained in five different clinics, using three different scanners, enabling the evaluation of different approaches in a realistic domain shift scenario.

In our slide-level experiments, we utilize tissue slides from The Cancer Genome Atlas (TCGA). The molecular characteristics of the dataset were evaluated in \citep{liu2018}, providing mutation status information such as Microsatellite Instability (MSI) for each sample. Considering the clinically relevant task of making predictions directly from whole tissue slides, we apply uncertainty estimation to the task of MSI prediction from WSIs in the TCGA colorectal dataset.

Our main contributions are:
\begin{itemize}
    \item A detailed analysis of the extent to which rejecting uncertain samples enhances classification performance on both tile and slide-level.
    \item A systematic comparison of the most prominent uncertainty estimation methods under domain shift in terms of selective classification, network calibration, and classifier performance on histopathological data.
    \item An investigation of the influence of label noise on the classification of WSIs and the robustness of the included uncertainty estimation methods against it.
	\item The release of an easily extendable code repository\footnote{\url{https://github.com/DBO-DKFZ/uncertainty-benchmark}} to facilitate further research on uncertainty estimation for deep neural networks.
\end{itemize}

\section{Methods}
\label{sec2:methods}

In this section, we describe the methods, uncertainty measures and evaluation settings used in our work. 

\subsection{Uncertainty Estimation Methods}

In uncertainty estimation, we want to compute the posterior predictive distribution of the output $\mb{y}$, given an input $\mb{x}$ and training data $\mc{D}$. 
This distribution can be formulated using Bayesian Model Averaging over the model's parameters $\mb{w}$ as 
\begin{align}
	p(\mb{y}|\mb{x}, \mc{D}) = \int p(\mb{y}| \mb{x}, \mb{w}) p(\mb{w} | \mc{D}) \,\mr{d} \mb{w} 
\end{align}

However, the posterior predictive distribution for neural networks is analytically intractable. As a result, in recent years several approximation methods have been proposed. Other methods are not motivated by Bayesian statistics but are nonetheless commonly used for quantifying predictive uncertainty. In the following, we briefly introduce the most prominent methods.
\\ \\
\textbf{Stochastic Variational Inference (SVI):}
\citep{blundell2015} and \citep{graves2011} approximate the posterior distribution by placing a Gaussian distribution over every parameter of the neural network. As an objective for the approximation, the estimated lower-bound (ELBO) is minimized \citep{blei2017}. We use the Flipout formulation \citep{wen2018a} of SVI for stabilizing the training procedure. 
\\ \\
\textbf{Monte-Carlo Dropout (MCDO):}
\citep{gal2016} show that the dropout operation, originally intended as a regularization method to stabilize neural network training, can be used to approximate the true posterior of the neural network. For the approximation, multiple forward passes of the same input, with activated dropout layers, are aggregated during inference. The distribution over the predictions obtained through this method can be seen as samples from an approximation of the posterior distribution. 
\\ \\
\textbf{Deep Ensemble:}
A Deep Ensemble \citep{lakshminarayanan2017} consists of multiple, in architecture identical, neural networks that are trained from different random initializations. The mean of all ensemble members serves as the prediction during inference. \citep{ovadia2019, ashukha2020} show that Deep Ensembles outperform many other methods in terms of calibration and robustness under domain shift. 
\\ \\
\textbf{Test-Time Data Augmentations (TTA):}
In contrast to Deep Ensembles, which employ multiple models during inference, TTA uses the same model multiple times by augmenting the input in different ways during inference. Generally, the same augmentations as at training time are applied during inference.
\citep{ayhan2018, ashukha2020} show the good performance of TTA in terms of robustness and calibration, which can come close to the performance of a Deep Ensemble while requiring less training time, as only one model is trained.
\\ \\
\textbf{Ensemble Variants:}
Additionally, ensemble variants of the methods SVI, MCDO and TTA are implemented and tested.

\subsection{Uncertainty Metrics}
\label{sec:uncertainty_metrics}
Given the predictions generated by the previously introduced methods, the literature developed multiple metrics to estimate the predictive uncertainty. 

The most commonly used approach in the field of calibration \citep{guo2017} is the so-called "confidence" of the prediction, which is the maximum of the softmax output. The core idea is that decisions that are far away from the decision boundary can be considered certain, while the most uncertain decisions lie close to the boundary, which is at $1/N$, where $N$ is the number of classes.
Other commonly used uncertainty metrics encompass the entropy  \citep{mobiny2019, band2021} of the prediction and the variance between predictions for ensemble-like outputs \citep{nair2020, leibig2017}. In \ref{apdx:uncertainty_metrics} we compare these approaches and conclude that the commonly used confidence measure is well-suited for this task.

\subsection{Evaluation Settings}
\label{sec:evaluation_settings}
This section covers the different evaluation settings and the performance metrics used in each setting.
We evaluate on tile-level with lesion-level annotations, as well as on the slide-level, where we only utilize slide-level labels.
\\ \\
\textbf{Reject Option:}
In a clinical setting, the model should be able to refer predictions with high uncertainty to human practitioners for evaluation, which is called selective classification \cite{geifman2017}. Models suitable for this task should assign higher uncertainties to their wrong predictions than to their correct predictions, thereby allowing to cut off a large number of false predictions, by thresholding at a certainty level. To compare multiple models on this ability, we compute the accuracy-reject curve \citep{nadeem2009}, plotting the achieved accuracy against the percentage of rejected data points in the dataset. We also compare the area under the curve of the accuracy-reject curves ($AUARC$). This evaluation has been recommended by Jaeger et al.\citep{jaeger2023}, where different evaluation practices in the field of failure detection have been compared and evaluated.
\\ \\
\textbf{Calibration:}
For measuring calibration, we utilize the Expected Calibration Error (ECE) \citep{guo2017, nixon2019}. Given a prediction for every data point in the dataset, the output probability or "confidence" for each sample should on average match the correctness of the prediction. In other words, we expect a prediction that has a confidence value of $60\%$ to be correct in about $60\%$ of the cases. To validate this intuition, the predictions are split into a predetermined number $M$ of bins $B$ of equal confidence range. Then the absolute difference between the average accuracy and confidence within each bin is summed up:
\begin{align}
    \label{eq:ece}
	\mr{ECE} = \sum_{m=1}^{M} \frac{|B_m|}{n} \Bigl| \mr{acc}(B_m) - \mr{conf}(B_m) \Bigl|
\end{align} 
Here, $|B_m|$ denotes the number of samples in the $m$-th bin and $n$ is the total number of samples.
\\ \\
\textbf{Label Noise:}
Medical annotations are often subject to an unquantified amount of label noise \citep{joskowicz2019, jensen2019, karimi2020a}, which may deteriorate the performance of supervised machine learning approaches. 
To our knowledge, the previously described methods have not been compared in their robustness to label noise in the medical domain. We evaluate the effect of label noise by creating multiple datasets with increasing levels and different types of label noise and evaluate the methods under these changing conditions.

\section{Experiment setup}

\subsection{Datasets and data processing}
\label{sec:dataset}

\textbf{Camelyon17}:
We conduct our experiments on the lesion-level annotated slides of the Cameylon17 dataset \citep{bandi2019}. This part of the Camelyon17 dataset consists of 50 WSIs of breast lymph node tissue, with annotated metastatic tissue. The slides were obtained from five different clinics in the Netherlands, using three different scanners, providing an ideal setting to assess the impact of domain shift between clinics on model performance. 

To induce a distribution shift between an in-distribution (ID) and an out-of-distribution (OOD) domain, we create two distinct splits of the contributing centers. The weak domain shift is based solely on location, as the out-of-distribtion (OOD) dataset contains scanners, that are also present in the in-distribution (ID) dataset. For this split, centers 0, 2, and 4 are part of the in-distribution (ID) dataset, containing the 3D Histech, Hamamatsu and Philips scanner, while centers 1 and 3 are used as OOD datasets, which both use the 3D Histech scanner.

To induce a strong domain shift, we split the centers in a manner, that the OOD datasets only contain scanners that are not present in the ID datasets, thereby creating an additional technological shift in the image acquisition process. 
For this, centers 0, 1 and 3 with the 3D Histech scanners are the ID datasets, with centers 2 and 4 being OOD (Philips and Hamamatsu).

For both splits, we then further partition the ID data into training, validation and test set. We sort the slides of each center based on the area of annotated tumor cells they contain and use the two median slides as the test set for each center. The training and validation sets are generated by a randomized $75\% / 25\%$ split of the tiles.

The tiles themselves are generated following \citep{khened2021} with median filtering and Otsu's thresholding of the HSV saturation component of the WSI image, followed by finally applying opening and closing dilation. After that, tiles of the size $256 \times 256$ are extracted. The tumor regions on the slide are indicated by polygonal annotations.
The annotations are used to compute the tumor coverage per tile. Tiles with more than $25\%$ tumor coverage are counted as tumor tiles and all tiles with $0\%$ tumor coverage are counted as non-tumor tiles. Tumor tiles with less than $25\%$ coverage by the tumor annotation are excluded from our standard training, to minimize the risk of label noise that could arise due to high inter-observer variability \citep{jensen2019, joskowicz2019, karimi2020a}. 
\\ \\
\textbf{TCGA}: 
In the slide-level analysis, we utilize tissue slides of colorectal cancer (CRC) from TCGA to predict the MSI status, following the approach described in \citep{kather2019, bilal2021}. Following the procedure outlined in \citep{kather2019}, we include only slides labeled as MSI-H and MSS, excluding slides with interfering markers or missing resolution information. This yields a total of 322 slides available for training and testing, comprising 59 MSI-labeled slides and 263 MSS-labeled slides. 
For the test set we define a domain shift by location, by allocating the 60 slides from the submitting centers MSKCC and Roswell Park for testing.

\subsection{Training setup and hyperparameter tuning}
\label{sec:training_setup}
For our tile-level experiments, we use a ResNet-34 \citep{he2016} with a batch size of 128 and a learning rate of 0.001 for all our experiments. As the optimizer, we utilize Adam \citep{kingma2017}, with a reduction of the learning rate by a factor of 10 if the validation loss, which is chosen as the cross-entropy loss, does not decrease for 3 epochs. For data augmentations, we follow \citep{tellez2019} applying random crops to size $224 \times 224$, random 90° rotations and color jitter (brightness:$\pm20\%$, contrast:$\pm30\%$, hue:$\pm10\%$, saturation:$\pm10\%$). The inputs are normalized with the mean and variance of the training data.
The best model is chosen by accuracy on the validation set. During training, we balance the training set by samples per class, but we do not balance the validation set, as we want to evaluate our methods on all available data.

For the Deep Ensemble architecture, we choose $n=5$ members following recent literature \citep{linmans2020, thagaard2020}. For MCDO, we place a dropout layer after each ResNet block, with a dropout probability $p=0.3$. This is in contrast to \citep{linmans2020, thagaard2020}, who only place a dropout layer before the last layer, observing no improvement in performance. For inference during testing, we use 10 SVI-, MCDO- and TTA- samples. Following \citep{wenzel2020}, we use an additional hyperparameter for weighting the influence of the SVI prior (Kullback-Leibler-Divergence to the normal distribution) on the training. 
We tune this hyperparameter as well as the dropout probability, the dropout layer placement and the learning rate with the python library Optuna \citep{akiba2019}.

For the slide-level experiments, we use the open-source CLAM method \citep{lu2021}. The CLAM pipeline consists of two central elements: First, tile-level features are extracted using a ResNet-18 \citep{he2016} pretrained on ImageNet \citep{deng2009}, such that each slide is represented as a bag of tile-level features. In the second step, an attention-based mechanism \citep{ilse2018} is applied to compute a slide-level prediction. For training the approach, we use the same setup as described in \citep{lu2021}, by combining the slide-level cross-entropy loss with a tile-level SVM clustering loss.
After extracting the test slides, we use $80\%$ of the remaining slides for training and $20\%$ for validation. We make sure that both labels MSI and MSS are distributed equally between all splits and use weighted sampling of both classes for training.
We set the initial learning rate to $2^{-4}$, monitor the validation loss with early stopping and set a minimum amount of $50$ and maximum amount of $200$ training epochs. The method is trained with dropout probability set to $p=0.25$. For our uncertainty evaluations, we train Deep Ensembles with $n=5$ members and for MCDO we use $10$ runs at inference.

All our experiments are conducted with PyTorch 1.11 on Nvidia GPUs with CUDA 11.3.

\begin{figure*}[ht]
	\centering
	\includegraphics[width=\textwidth]{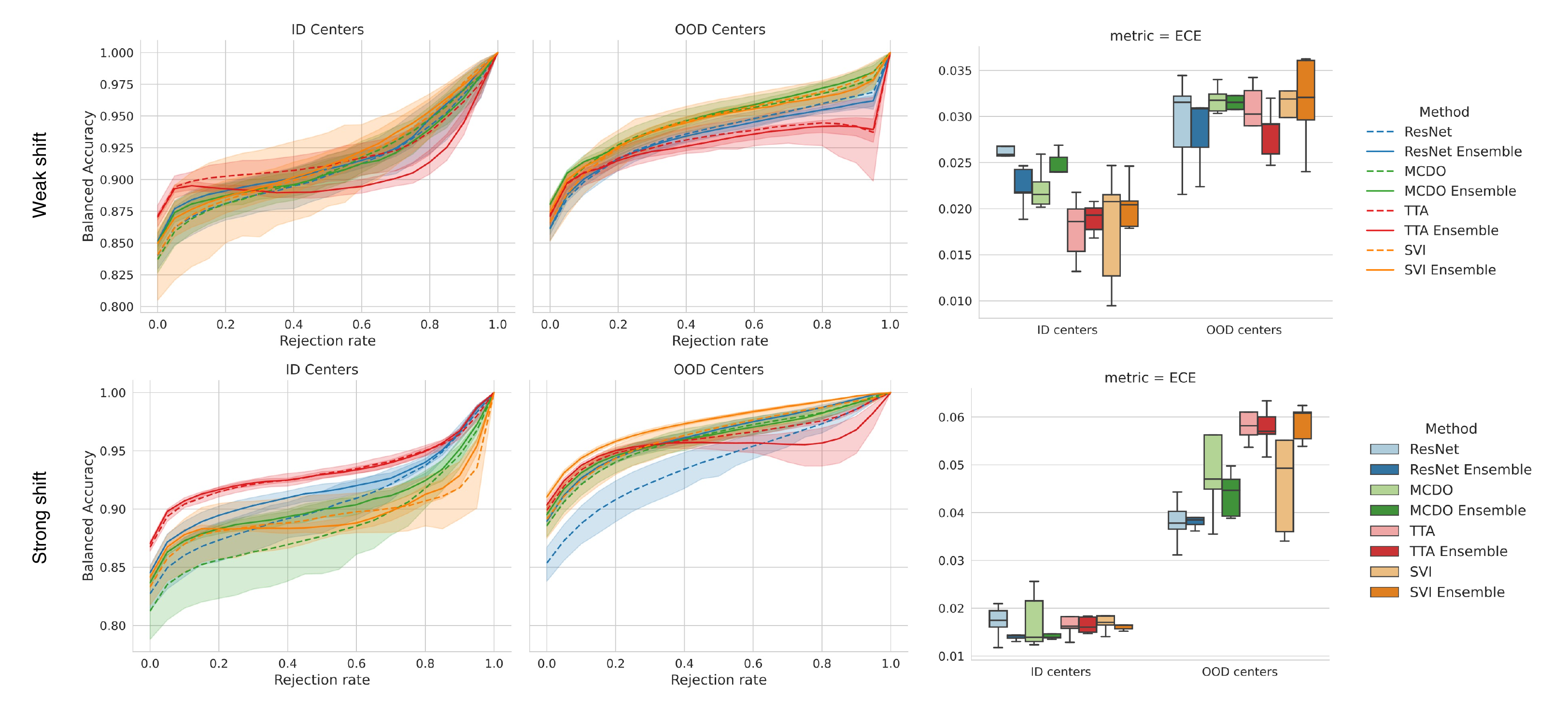}
	\caption{(Balanced) Accuracy-Reject Curves on ID and OOD centers, as well as the calibration error, for the weak and strong domain shift setup. The x-axis contains the proportion of rejected inputs over the dataset, which is plotted against the reached Balanced Accuracy on the remaining images. The order of the excluded points was determined by the predictive confidence (see \ref{apdx:uncertainty_metrics}). We plot the mean curves over all 5 trials, with the shaded area denoting the $95\%$ confidence interval. In the plots on the right side, we compare the influence of the uncertainty estimation methods on network calibration in terms of median ECE over all slides of the test sets, on both ID and OOD centers.}
	\label{fig:reject_vs_acc}
\end{figure*}

\begin{table}[tbh]    
    \caption{Mean and standard deviation of the area under the curve of the accuracy-reject curves ($AUARC$) presented in \autoref{fig:reject_vs_acc}.}
    \begin{tabular}{l|l|cc} \label{tab:AUC-MRC}
    Shift             & Method             &             ID Centers &            OOD Centers \\
    \hline
    \multirow{8}{*}{Weak} & ResNet &  $ 0.9105_{\pm0.0190}$ &  $ 0.9375_{\pm0.0094}$ \\
                 & ResNet Ensemble &  $ 0.9169_{\pm0.0071}$ &  $ 0.9354_{\pm0.0044}$ \\
                 & MCDO &  $ 0.9124_{\pm0.0134}$ &  $ 0.9470_{\pm0.0112}$ \\
                 & MCDO Ensemble &  $ 0.9126_{\pm0.0104}$ &  $ \mb{0.9494}_{\pm0.0034}$ \\
                 & TTA &  $ \mb{0.9200}_{\pm0.0140}$ &  $ 0.9305_{\pm0.0169}$ \\
                 & TTA Ensemble &  $ 0.9051_{\pm0.0053}$ &  $ 0.9283_{\pm0.0056}$ \\
                 & SVI &  $ 0.9156_{\pm0.0329}$ &  $ 0.9467_{\pm0.0148}$ \\
                 & SVI Ensemble &  $ 0.9183_{\pm0.0133}$ &  $ 0.9459_{\pm0.0051}$ \\
    \hline
    \multirow{8}{*}{Strong} & ResNet &  $ 0.9057_{\pm0.0169}$ &  $ 0.9402_{\pm0.0150}$ \\
                 & ResNet Ensemble &  $ 0.9184_{\pm0.0070}$ &  $ 0.9639_{\pm0.0043}$ \\
                 & MCDO &  $ 0.8864_{\pm0.0303}$ &  $ 0.9609_{\pm0.0093}$ \\
                 & MCDO Ensemble &  $ 0.9048_{\pm0.0075}$ &  $ 0.9625_{\pm0.0070}$ \\
                 & TTA &  $ 0.9321_{\pm0.0047}$ &  $ 0.9607_{\pm0.0024}$ \\
                 & TTA Ensemble &  $ \mb{0.9332}_{\pm0.0024}$ &  $ 0.9547_{\pm0.0095}$ \\
                 & SVI &  $ 0.8948_{\pm0.0269}$ &  $ 0.9643_{\pm0.0110}$ \\
                 & SVI Ensemble &  $ 0.8958_{\pm0.0058}$ &  $ \mb{0.9739}_{\pm0.0015}$ \\
    \hline
    \end{tabular}
\end{table}

\begin{figure*}[t]
	\centering
	\includegraphics[width=\textwidth]{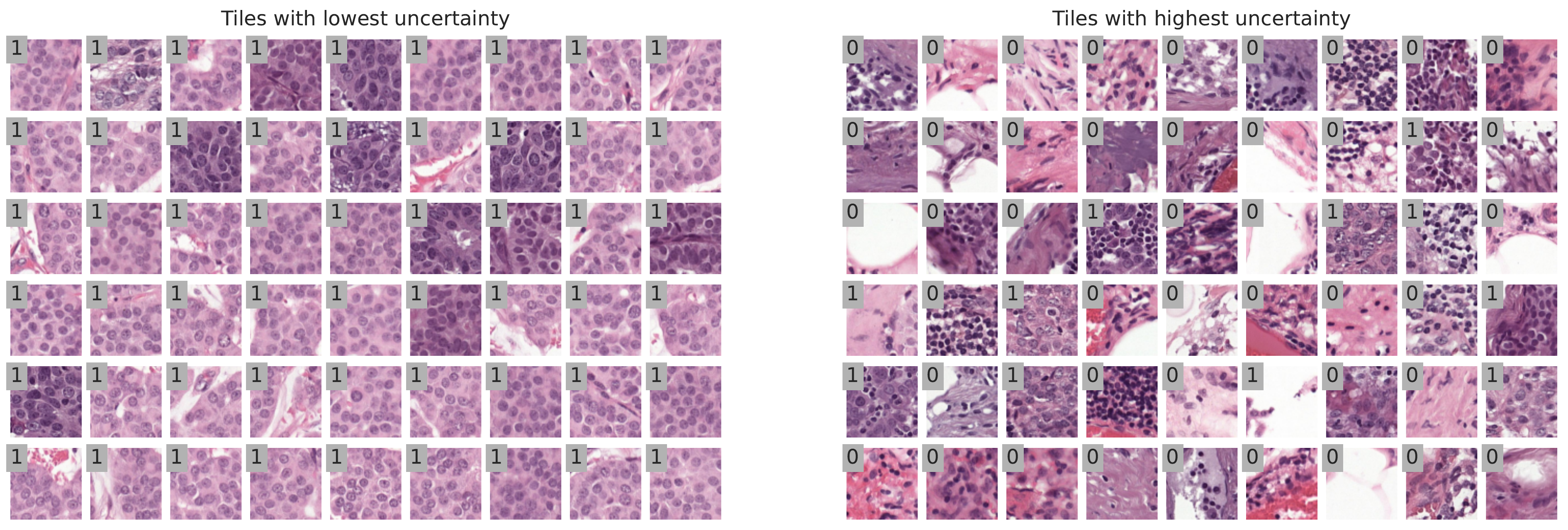}
	\caption{Least and most uncertain tiles from the in-distribution test set computed by the confidence of a ResNet Ensemble. The numbers represent the ground-truth labels: '1' indicates tumor and '0' indicates a non-tumor tile.}
	\label{fig:tiles_by_uncertainty}
\end{figure*}

\section{Results} 
\label{sec:Results}
%\the\textwidth 522pt // 7.25 inches
%\the\linewidth 252.25pt // 3.5 inches
%\makeatletter
%    \f@size % Fontsize is 10pt
%\makeatother

We trained each method 5 times with different seeds, reporting their average or median performance.

\subsection{Reject Option}
\label{sec:reject_option}

We compare the uncertainty methods in terms of their ability to detect mispredictions. For that, we compute the (Balanced) Accuracy-Reject Curves (see Section \ref{sec:evaluation_settings}).
In \autoref{fig:reject_vs_acc}, we compare the methods and their performance with an increasing ratio of rejected tiles, for the test slides on the ID centers and OOD centers on the weak and strong domain shift, respectively. To determine the most uncertain tiles that are rejected, we use the confidence measure.
On the ID centers of the weak shift setup, all methods perform similarly well, with the exception being the TTA Ensemble, which starts with higher balanced accuracy, but is the only method that does not show a monotonic increase of balanced accuracy with a growing rejection rate. 
A similar trend can be observed on the OOD centers, where TTA shows stagnating performance after rejecting more than $50\%$ of uncertain tiles. 

In contrast, within the strong shift setup, TTA and TTA Ensemble achieve the highest balanced accuracies along the entire curve on the ID centers, while the MCDO and SVI methods perform worst. On the OOD centers the ranking is mostly reversed, with the SVI Ensemble leading to the highest accuracy-reject-curves, while the baseline ResNet and the TTA methods perform worst.
Surprisingly, the Balanced Accuracy values on the OOD centers are partially higher than on the ID centers. 
In \ref{apdx:metric_choices_detailed_results}, we elaborate on this behavior on the OOD data. Additionally, we evaluate and compare the classification performance of these methods using the median balanced accuracy over all slides of the test datasets.

In \autoref{tab:AUC-MRC} we show the area under the curve for the accuracy-reject curves ($AUARC$) for the weak and strong domain shift scenarios.
For the ID centers of the weak domain shift, all methods but TTA Ensemble improve upon the $AUARC$ compared to the baseline ResNet, with TTA, SVI Ensemble and ResNet Ensemble performing the best.
On the in-distribution data of the strong shift, however, all SVI and MCDO methods perform worse or are comparable to the ResNet baseline. Here both TTA approaches and the ResNet Ensemble are the only methods to substantially improve upon the ResNet baseline. 
By using the $AUARC$ metric, we can identify the MCDO Ensemble as best performing method on the OOD centers of the weak shift setup and the SVI Ensemble as best performing method on the OOD centers of the strong shift setup, while the TTA methods underperform on the OOD data of the weak shift. On the OOD data of the strong shift, every method improves upon the baseline.

From our experiments, no single method can be identified, that performs best between all data splits, as the rankings of the methods vary heavily between the ID and OOD data, as well as between the weak shift and strong shift setup. However, ensemble approaches often outperformed their singular counterparts under the domain shift scenarios. Further analysis of the ranking of the uncertainty methods in a leave-one-out setup by centers, which comes to the same conclusion,  can be found in \ref{apdx:camelyon-leave-out}.

In \autoref{fig:tiles_by_uncertainty} we present a collection of the most certain and most uncertain tiles within the ID test data of the strong shift setup, computed with a ResNet Ensemble. 
We observe that the neural network appears to be most confident on tumor tiles (label 1), that cover the whole tile and possess a similar cell structure. For the most uncertain tiles on the right side, no comparable structure among the tiles is observable. These tiles both contain tumor and non-tumor tissue and often lie at border regions between different tissue types.  
Many uncertain tiles seem to lie at the border of annotated tumor regions, that we suspect to have a larger degree of label noise and are harder to distinguish due to features that are present in both healthy and tumorous tissue.

\subsection{Calibration}

In the plot on the right-hand side of \autoref{fig:reject_vs_acc}, we evaluate model calibration in terms of ECE (see Section \ref{sec:evaluation_settings}) for the weak and strong shift. 
The ECE values have been computed as the median calibration error over all slides, as explained in \ref{apdx:metric_choices_detailed_results}.
On the ID centers in the weak shift setup, the TTA and SVI methods perform particularly well. On the ID centers for the strong shift setup, the best-performing methods are the ResNet Ensemble and MCDO Ensemble, while the TTA and SVI methods perform worse. 

Under distribution shift, the best method is in fact the ResNet baseline method and its ensemble variant. All other methods perform worse on both weak and strong shift settings, with TTA Ensemble performing still quite well under the weak shift, but worst in the strong shift setting, along with the SVI Ensemble. 

In conclusion like with other metrics presented, we cannot identify a best-performing method. Only ensembling approaches nearly consistently outperform their singular counterparts. 
\autoref{tab:BoxplotDetailedResults} contains the detailed numeric results for our calibration experiments.

\subsection{Label Noise}
\label{sec:results_label_noise}

\begin{figure*}[tbh]
	\centering
	\includegraphics[width=\textwidth]{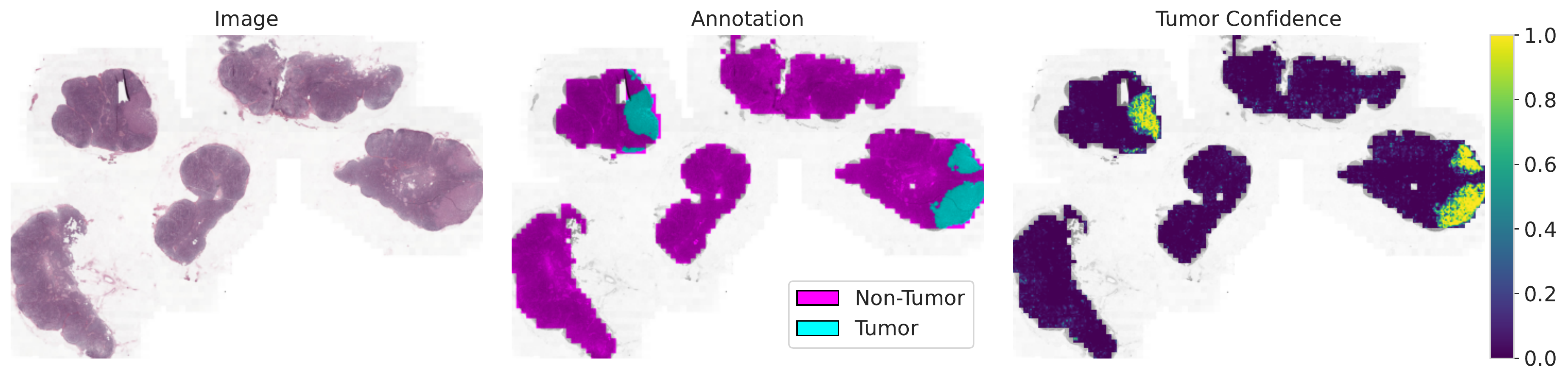}
	\caption{WSI (\textit{patient\_017\_node\_2}) with ground-truth annotations and model predictions. Predictions have been generated from a ResNet Ensemble. The tumor confidence decreases in areas near the border of the annotation, while the uncertainty thereby increases. This result is consistent over the whole dataset.}
	\label{fig:slide_vis}
\end{figure*}

\begin{figure*}[tbh]
	\centering
	\includegraphics[width=\textwidth]{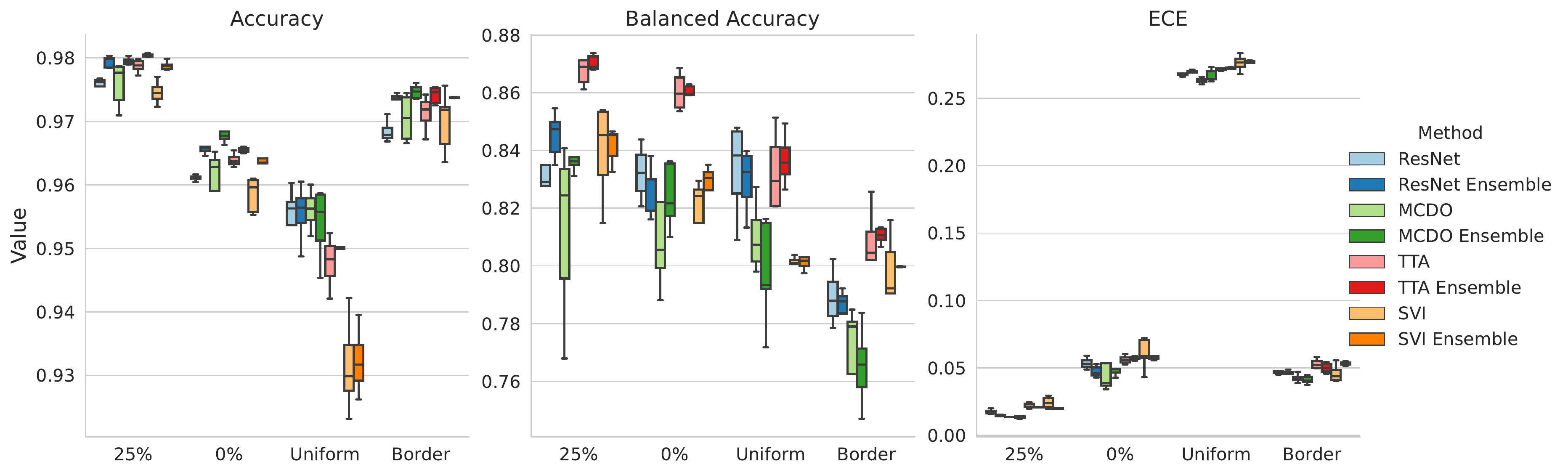}
	\caption{Performance of the proposed methods on different datasets under label noise. We compare the performance on the original dataset ($25\%$), a dataset with a $0\%$ tumor coverage threshold for tumor tiles ($0\%$), applying uniform noise to the tile labels (Uniform) and randomly flipping labels of tiles which are located at the border of the annotation (Border).}
	\label{fig:LabelNoiseBoxPlot}
\end{figure*}

For translating our tile-level observations to slide-level, we stitch the tile-level predictions back to a tumor confidence map on slide-level that we show on the example of one Camelyon17 slide in \autoref{fig:slide_vis}. 
When observing the generated confidence maps, we can see lower tumor confidence at the border of annotated tumor regions. 

Building on these observations, in our label noise experiments, we investigate the effects of imprecise tumor annotations in the border area of annotated tumor regions. To this end, we define three supplementary training datasets building on the in-distribution Camelyon17 dataset of the strong shift scenario by introducing different types of label noise to the annotations.
We first create a dataset by setting the inclusion threshold by tumor coverage for tumor tiles to $0\%$. We previously excluded every tile that was covered by less than $25\%$ by tumor annotations (see Section \ref{sec:dataset}) to reduce the chance of label noise. The other two datasets are created by applying random label noise to the training split of the $0\%$ threshold dataset. First, we apply uniform label noise to the whole slide, with a $25\%$ chance of flipping the tile class.
As this type of annotation noise does not reflect real-world inter-observer variability, we next apply label flipping to the border regions of the annotation. We flip the labels of the tumor tiles, which lie at the border of the annotation polygon and thereby are not fully covered by annotated tumor cells. We set the chance of this event occurring to $25\%$ per tile. 

In \autoref{fig:LabelNoiseBoxPlot} we show the results of the label noise experiments. Detailed results can be found in \autoref{tab:LabelNoiseDetailedResults}. All methods were trained on the noisy datasets but evaluated on the original test dataset.
In terms of accuracy, the ensemble methods outperform their non-ensemble counterparts by a significant margin. Only on the Uniform dataset, ResNet and MCDO perform similarly to their ensemble counterparts ($\sim 95.6\%$ accuracy for each method). 
When viewing the balanced accuracy metric, TTA and TTA Ensemble exceed every other method by a large margin. Ensemble methods again consistently outperform their singular counterparts. The MCDO and SVI methods are the least robust methods when exposed to label noise, often performing worse than the single ResNet baseline.

We can conclude that ensembling approaches are not only more robust to domain shifts and image corruptions \citep{ovadia2019} but in a similar manner also to label noise, in our case in histopathological images. From our experiments, SVI and MCDO, however, are not fit to deal with label noise often leading to only slightly improved or even worse results. 

TTA however does not perform well in terms of calibration error. Here MCDO outperforms TTA and SVI, which produced the overall worst calibrated predictions, in contrast to recent literature \citep{ashukha2020, ayhan2020}. We can see a large increase in ECE on the dataset with uniform label noise and a slight increase in the miscalibration on the other two datasets with label noise compared to our baseline dataset. Except for the original dataset, no trend of ensembling methods decreasing calibration error is visible. Ensembling does not seem to improve calibration when confronted with larger quantities of label noise, contrary to the setting of domain shift (\autoref{fig:reject_vs_acc}) where ensembling decreased the calibration error.

\subsection{Slide-level Analysis}
\label{sec:results_slide_level}

For the slide-level analysis, we aim to predict the MSI status on WSIs 
of colorectal cancer tissue from the TCGA dataset as described in Section \ref{sec:dataset}. Since we evaluated the performance of common uncertainty methods on the tile-level before, we first try to infer a slide-level prediction by using tile-level uncertainties. A similar approach has been presented in the concurrent work by \citep{linmans2023} for the slide-level prediction of prostate cancer.
To train the tile-level approach, at first we use a subtyper network to identify tumor tiles on each WSI. We then assign the binary slide-level label (MSS or MSI) to each tumor tile of the corresponding WSI and train an ensemble of five ResNet-34, comparable to the previous tile-level evaluations. To compute a slide-level prediction, we average the predictions of the top 1\% most confident tile-level scores. By evaluating five different runs using the described method, we retrieve AUROC values with mean and standard deviation $0.6243 \pm 0.0610$ on the test set. 

Besides aggregating tile-level predictions, there exist more sophisticated methods in the context of WSI classification, one of which is the CLAM method \citep{lu2021}. We trained and evaluated the CLAM method on the same data splits and the AUROC values over 5 runs are $0.7168 \pm 0.0328$. The significant increase in performance leads to the conclusion that the attention-based CLAM approach is better suited for the slide-level prediction task than aggregating tile-level predictions by uncertainty. 

\begin{figure}[tbh]
	\centering
	\includegraphics[width=0.5\linewidth]{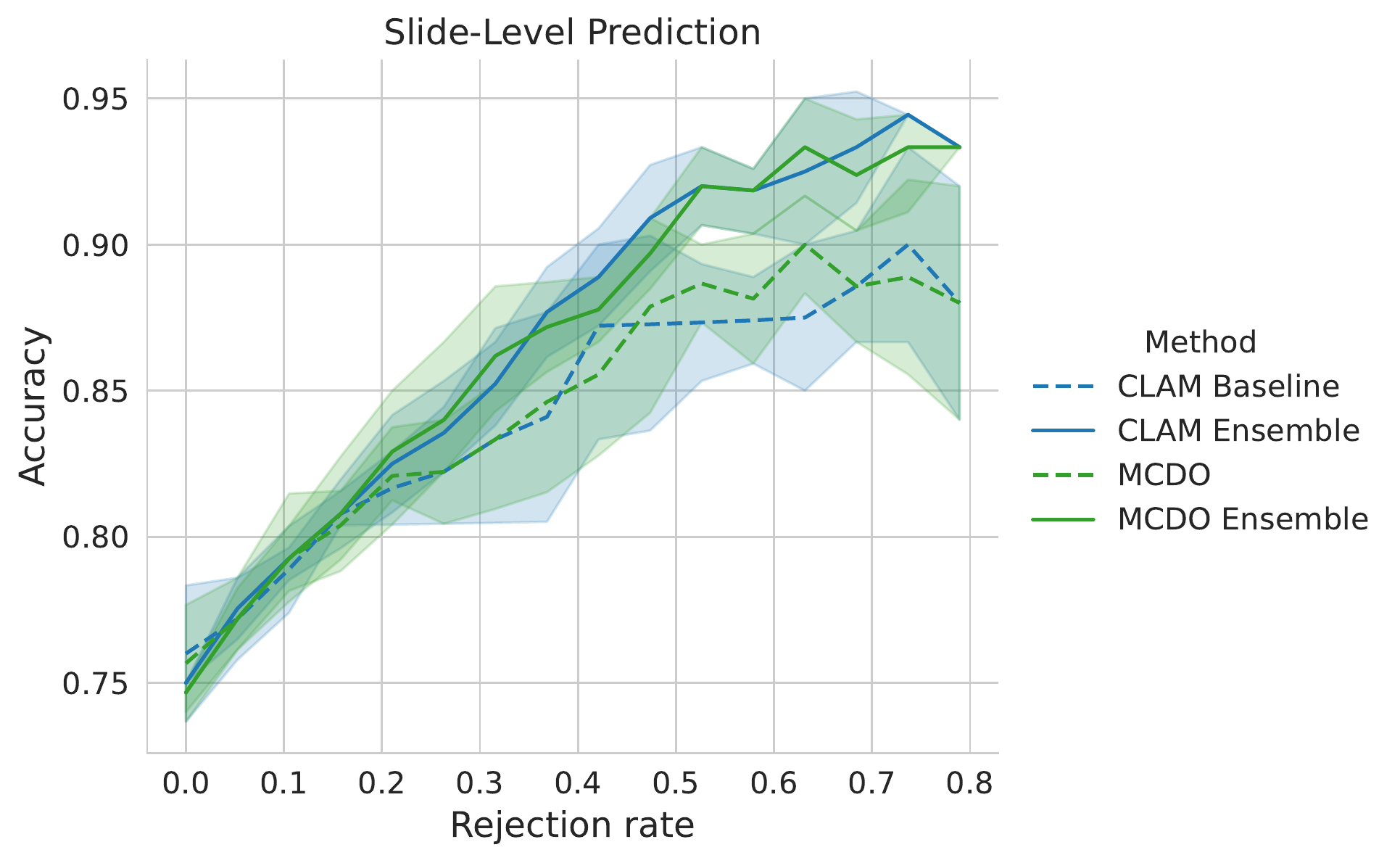}
	\caption{Accuracy-Reject Curves for slide-level MSI prediction using the CLAM method \citep{lu2021}}
	\label{fig:clam_acc_reject}
\end{figure}

Following our tile-level experiments, we evaluate the ability for selective classification on the slide-level under domain shift. For this, we implemented Deep Ensembles and Monte-Carlo Dropout on top of the better performing CLAM-method. In \autoref{fig:clam_acc_reject} we show the Accuracy-Reject curves for the slide-level prediction tasks. From the curves, we can see that rejecting slides by uncertainty leads to improvements in performance, comparable to the tile-level experiments. We also observe a similar trend as in the tile-level evaluations, that ensembles of methods perform better than their single model counterparts. 

Overall, we can conclude that rejecting slide-level predictions by uncertainty is a viable approach to selective classification and can significantly increase predictive accuracy.

\section{Discussion}

In the previous section, we have gone through an extensive comparison of the most prominent methods for uncertainty estimation under domain shift on histopathological WSIs.
In this section, we want to discuss the observations that we have made and we want to formulate recommendations for other researchers that try to integrate uncertainty estimation into digital pathology. 

Our results show that mispredictions can be detected reliably and that the right methods can increase the robustness to domain shift and label noise, while also providing better-calibrated predictions.

Among the methods for uncertainty estimation, ensembles lead to the most reliable uncertainty estimates and additionally improve classification performance and network calibration. 
As expected, combining MCDO, TTA and SVI with ensembling can lead to further improvements in classification performance, however, it also entails a steep increase in computational requirements, which might not be possible in some medical environments.
In the case of MCDO and SVI, architectural changes in the neural network need to be made and these methods performed the most inconsistent across our experiments. As the rankings of all investigated methods varied heavily between different settings, see also \ref{apdx:camelyon-leave-out}, no best method can be identified for all tasks.

In terms of misprediction detection on ID data, all methods provide a reliable improvement in classification performance, with a growing rejection rate making misprediction detection a feasible scenario in a clinical setting. The TTA methods performed the most inconsistent for selective classification, as they were the only methods with a non-monotonic increase of the accuracy-reject curves. 
The largest increase for every method is at the beginning of the accuracy-reject-curve, with the slope often decreasing notably around the $20\%$ reject rate threshold. From our results, it is plausible that in many cases it could be enough to only reject around $20\%$ of most uncertain predictions to receive a significant accuracy boost. 

We could observe a strong increase in calibration error from the ID centers to the OOD centers. However, from our experiments, no single non-ensemble method outperformed the other approaches over all domain shifts and data splits in terms of calibration error. Ensembles often improved the measured calibration.

By sorting the tile predictions by uncertainty, we observe visually recognizable differences between the most certain and most uncertain tiles, which are consistent with all methods. On the WSIs, the networks make especially confident predictions for tiles that lie inside tumor metastases, while being the most uncertain at the border of tumor regions, which can also be seen when visualizing the uncertainty on the slide-level.
In our label noise experiments, we could immediately see the impact of including tissue-border tiles with noisy labels, with a drop in classifier performance. Care should therefore be taken in the correct preparation of the training data, as inaccurate annotations or label noise induced by the tiling process in the border regions can significantly impact performance.
In our experiments, the ensemble approaches and TTA showed increased resistance to label noise.

Besides analyzing the influence of uncertainty estimation methods on the tile-level tumor prediction task, we have also applied uncertainty estimation to the slide-level task of MSI prediction on slides of colorectal cancer tissue. Since clinical diagnoses are performed on slide-level, it is an important insight that with the help of uncertainty estimation, mispredictions can be identified on slide-level and the performance of the neural network can be improved. Here ensembles again performed better than their singular counterparts, while MCDO did not improve upon the baseline CLAM method.

The correct choice of metrics is an important consideration when dealing with histopathological data. In our experiments, we noticed that different splits of the dataset can lead to considerably  different results, see \ref{apdx:camelyon-leave-out}. Evaluation therefore requires robust metrics. In our tile-level experiments we chose to use the median performance of the metric over all slides in the test dataset which proved to be more robust, see \ref{apdx:metric_choices_detailed_results}.

%To extend our work, additional methods could be compared, for example, integrating multi-head ensembles \citep{linmans2020} or deterministic uncertainty methods \citep{postels2022a}, which offer a sampling-free and fast alternative. Our label noise experiments could be extended to more methods and more realistic scenarios.

\section{Conclusion}
Deployment of AI-based diagnostic systems in the safety-critical area of histopathology demand uncertainty-aware machine learning algorithms, which generate trust in the model’s predictions.

To this end, we compared multiple uncertainty estimation methods and uncertainty metrics across domain shift and label noise scenarios in their performance, calibration and ability to detect mispredictions in the histopathological setting on tile- and slide-level.
Our results show that all investigated methods are well-capable to detect mispredictions and reject inputs they are unlikely to classify correctly, which is of high clinical relevance for the reliable and safe deployment of machine learning systems. In our experiments TTA is the only method that does not lead to a monotonic increase in performance with a growing ratio of rejected inputs. Furthermore, ensembling approaches improve the calibration on histopathological data over their singular counterparts on ID and OOD data, while on the other hand, none of the included robustness and uncertainty estimation methods from the classic computer vision literature could reliably improve the calibration under domain shift.

Label noise can have a substantial effect on the quality of the predictions. As label noise can be a common problem in medical data, great care should be put into identifying potentially mislabeled inputs and in choosing data preparation techniques that produce well-labeled data points. In terms of robustness to label noise, ensembles and TTA approaches performed best in our experiments, while no method proved robust to the strong drop in calibration under label noise.

In general, ensemble approaches performed the most consistent over all our experiments.
Contrary to results in the classical computer vision literature MCDO and SVI often only performed on par or slightly better than the baseline ResNet on most measures, but require substantially more computational resources during inference and architectural changes to the neural network architecture. TTA produced the best classification results on ID data, however, this advantage did not carry over under our domain shift scenarios. Additionally, it produced the worst-behaved accuracy-reject curves for the selective classification task.
Overall, the ranking between the methods is inconsistent and we can not recommend one singular best-performing approach for all scenarios. As ensembling approaches lead to relatively consistent improvements over their singular counterparts they can from our experience be recommended, while also being comparatively easy to implement and use in practice. 

While no single benchmark can give all-encompassing results and insights, we hope that our evaluation gives guidance for the utilization of uncertainty methods in the area of histopathology. Our published code is designed to be easily reproducible and extendable to further studies. 

\section*{Acknowledgments}

The research is funded by the \textit{Ministerium für Soziales und Integration}, Baden Württemberg, Germany.

\newpage

\typeout{}
\bibliography{refs}

\onecolumn
\appendix
\section{Metric Choices and Detailed Results}
\label{apdx:metric_choices_detailed_results}

In this section, we motivate the usage of our metrics and give detailed results for all our experiments.

Histopathological datasets commonly exhibit a high degree of class imbalance, with classes like healthy tissue being over-represented compared to classes like tumor tissue. As a result, commonly used metrics such as AUROC and accuracy may yield misleading results, as they are highly sensitive to class imbalances. Furthermore, there can be a significant variation in the distribution of classes from slide to slide, with one slide containing minimal tumor cells while the next slide predominantly consists of a large metastasis.

In our experiments, this is the reason why we can observe higher balanced accuracy values on the OOD centers in \autoref{fig:reject_vs_acc}. The OOD center 3 for the weak shift and 4 for the strong shift both contain slides with an unusual amount of tumor tissue, which results in many tiles, where the tumor tissue covers the whole tile, which as seen in \autoref{fig:tiles_by_uncertainty} are the tiles where the neural networks are the most confident in their predictions. This skew in the distribution of tumor tiles, towards tiles with 100\% tumor coverage and fewer border cases, in turn, leads to higher balanced accuracy values on the OOD centers. This can also be seen in the higher $AUARC$ values on center 4 compared to center 2 under the strong shift scenario in \autoref{tab:comp-uncer-metrics}.

Such problems make reliably estimating the performance of a classifier a difficult task, as different splits of the dataset can result in very different slides for the test dataset, which in our experiments had large impacts on the measured scores and rankings of the methods. 

For this reason, we report the median of our measured metrics over all slides of the test dataset. While in our experiments there were always outlier slides where metrics performed above or below average, our approach proved to be more stable to the choice of dataset split.

However, we did not employ this approach for the accuracy-reject curves of \autoref{fig:reject_vs_acc} and \autoref{tab:AUC-MRC}, as interpreting the median curve of multiple accuracy-reject curves can be challenging. The primary focus of this task is to examine the behavior of the measured metric curve under tile-level rejection and the relative rankings of the methods. In the following, you can find the median balanced accuracy values over all slides of the test dataset.

\autoref{tab:BoxplotDetailedResults} shows the detailed results of the classifier performance of our Camelyon17 experiments, with the median balanced accuracy and ECE over all slides of the test dataset reported.

On the ID data under both domain shifts, we can see each ensemble method improving upon its singular counterpart with only TTA Ensemble under the weak shift being on par. However, by themselves, MCDO as well as SVI, perform worse than the ResNet baseline, while TTA slightly outperforms the baseline. 

Under the weak and strong domain shifts none of the alternative non-ensemble approaches improves upon the ResNet baseline, which is contrary to other reports outside of the field of histopathology \citep{ovadia2019}. Most ensemble approaches however improve upon the ResNet baseline and consistently perform better than their singular counterparts.

Similar to our discussion in \autoref{sec:Results}, no best method can be deduced from our experimental results. Still, ensemble approaches outperform baseline methods and consistently can be recommended.

\autoref{tab:LabelNoiseDetailedResults} shows the detailed results of our label noise experiments.

\begin{table*}[tbh]
\centering
\caption{Detailed results for tile classifier performance on Camelyon17. We report the median value over 5 runs and the interquartile range for each domain shift, data split and method. The metrics are the median performance over all slides of the test dataset.}
\label{tab:BoxplotDetailedResults}

\begin{tabular}{l|ll|cc}
\hline 
        Shift &       Split &             Method & $\text{Balanced Accuracy} \uparrow$ &  $\text{ECE} \downarrow$ \\
\hline \hline

\multirow{16}{0.14\textwidth}{Weak} & \multirow{8}{0.14\textwidth}{ID Centers}
           & ResNet &  $ 0.8551_{\;0.0264}$ &  $ 0.0259_{\;0.0010}$ \\
           & & ResNet Ensemble &  $ 0.8712_{\;0.0014}$ &  $ 0.0218_{\;0.0025}$ \\

           & & MCDO &  $ 0.8542_{\;0.0356}$ &  $ 0.0215_{\;0.0024}$ \\
           & & MCDO Ensemble &  $ 0.8691_{\;0.0054}$ &  $ 0.0240_{\;0.0016}$ \\

           & & TTA &  $ 0.8689_{\;0.0058}$ &  $ 0.0186_{\;0.0046}$ \\
           & & TTA Ensemble &  $ 0.8669_{\;0.0022}$ &  $ 0.0193_{\;0.0023}$ \\
           
           & & SVI &  $ 0.8321_{\;0.0368}$ &  $ 0.0208_{\;0.0088}$ \\
           & & SVI Ensemble &  $ 0.8731_{\;0.0194}$ &  $ 0.0204_{\;0.0027}$ \\

\cline{2-5}

&\multirow{8}{0.14\textwidth}{OOD Centers} 

           & ResNet &  $ 0.6961_{\;0.0245}$ &  $ 0.0315_{\;0.0055}$ \\
           & & ResNet Ensemble &  $ 0.6863_{\;0.0130}$ &  $ 0.0308_{\;0.0043}$ \\
           
           & & MCDO &  $ 0.6842_{\;0.0447}$ &  $ 0.0318_{\;0.0018}$ \\
           & & MCDO Ensemble &  $ 0.6944_{\;0.0109}$ &  $ 0.0315_{\;0.0015}$ \\

           & & TTA &  $ 0.6871_{\;0.0324}$ &  $ 0.0303_{\;0.0038}$ \\
           & & TTA Ensemble &  $ 0.7046_{\;0.0100}$ &  $ 0.0292_{\;0.0033}$ \\

           & & SVI &  $ 0.6732_{\;0.0346}$ &  $ 0.0319_{\;0.0029}$ \\
           & & SVI Ensemble &  $ 0.6860_{\;0.0113}$ &  $ 0.0321_{\;0.0065}$ \\

\hline

\multirow{16}{0.14\textwidth}{Strong} & \multirow{8}{0.14\textwidth}{ID Centers} 
           &              ResNet &  $ 0.8019_{\;0.0031}$ &  $ 0.0175_{\;0.0034}$ \\
           &             & ResNet Ensemble &  $ 0.8131_{\;0.0014}$ &  $ 0.0142_{\;0.0006}$ \\
        & & MCDO &  $ 0.7952_{\;0.0214}$ &  $ 0.0139_{\;0.0085}$ \\
           &             & MCDO Ensemble &  $ 0.8140_{\;0.0086}$ &  $ 0.0139_{\;0.0008}$ \\

           &             & TTA &  $ 0.8105_{\;0.0098}$ &  $ 0.0162_{\;0.0024}$ \\
           &             & TTA Ensemble &  $ 0.8222_{\;0.0080}$ &  $ 0.0161_{\;0.0032}$ \\

           &             & SVI &  $ 0.7907_{\;0.0104}$ &  $ 0.0171_{\;0.0019}$ \\
           &             & SVI Ensemble &  $ 0.8126_{\;0.0074}$ &  $ 0.0165_{\;0.0008}$ \\

\cline{2-5}
&\multirow{8}{0.14\textwidth}{OOD Centers} 
           &              ResNet &  $ 0.7402_{\;0.0735}$ &  $ 0.0378_{\;0.0037}$ \\
           &             & ResNet Ensemble &  $ 0.7940_{\;0.0005}$ &  $ 0.0385_{\;0.0014}$ \\
        & & MCDO &  $ 0.7351_{\;0.0250}$ &  $ 0.0470_{\;0.0113}$ \\
           &             & MCDO Ensemble &  $ 0.7662_{\;0.0095}$ &  $ 0.0447_{\;0.0077}$ \\

           &             & TTA &  $ 0.7317_{\;0.0232}$ &  $ 0.0582_{\;0.0048}$ \\
           &             & TTA Ensemble &  $ 0.7348_{\;0.0295}$ &  $ 0.0570_{\;0.0036}$ \\

           &             & SVI &  $ 0.6962_{\;0.0554}$ &  $ 0.0493_{\;0.0191}$ \\
           &             & SVI Ensemble &  $ 0.7486_{\;0.0119}$ &  $ 0.0608_{\;0.0056}$ \\

\hline
\end{tabular}
\end{table*}

\begin{table*}
\centering
\caption{Detailed results for our experiments under label noise. We report the median and the interquartile range as shown in the boxplots.}
\label{tab:LabelNoiseDetailedResults}

\begin{tabular}{ll|ccc}
\hline
      Experiment  &        Method      &             $\text{Accuracy} \uparrow$  &      $\text{Balanced Accuracy} \uparrow$ &  $\text{ECE} \downarrow$ \\
\hline \hline
\multirow{8}{0.14\textwidth}{25\%}
        & ResNet &  $ 0.9762_{\;0.0010}$ &  $ 0.8291_{\;0.0073}$ &  $ 0.0181_{\;0.0020}$ \\
        & ResNet Ensemble &  $ 0.9798_{\;0.0015}$ &  $ 0.8474_{\;0.0105}$ &  $ 0.0145_{\;0.0011}$ \\
        & MCDO &  $ 0.9777_{\;0.0053}$ &  $ 0.8244_{\;0.0380}$ &  $ \mb{0.0135}_{\;0.0001}$ \\
        & MCDO Ensemble &  $ 0.9794_{\;0.0006}$ &  $ 0.8363_{\;0.0026}$ &  $ 0.0137_{\;0.0010}$ \\
        & TTA &  $ 0.9788_{\;0.0013}$ &  $ 0.8689_{\;0.0076}$ &  $ 0.0214_{\;0.0026}$ \\
        & TTA Ensemble &  $ \mb{0.9805}_{\;0.0003}$ &  $ \mb{0.8690}_{\;0.0043}$ &  $ 0.0207_{\;0.0003}$ \\
        & SVI &  $ 0.9745_{\;0.0019}$ &  $ 0.8453_{\;0.0219}$ &  $ 0.0240_{\;0.0066}$ \\
        & SVI Ensemble &  $ 0.9787_{\;0.0007}$ &  $ 0.8452_{\;0.0076}$ &  $ 0.0196_{\;0.0007}$ \\
\hline
\multirow{8}{0.14\textwidth}{0\%}
        & ResNet &  $ 0.9612_{\;0.0004}$ &  $ 0.8323_{\;0.0124}$ &  $ 0.0531_{\;0.0046}$ \\
        & ResNet Ensemble &  $ 0.9658_{\;0.0007}$ &  $ 0.8299_{\;0.0110}$ &  $ 0.0460_{\;0.0061}$ \\
        & MCDO &  $ 0.9628_{\;0.0048}$ &  $ 0.8055_{\;0.0229}$ &  $ \mb{0.0387}_{\;0.0165}$ \\
        & MCDO Ensemble &  $ \mb{0.9677}_{\;0.0012}$ &  $ 0.8217_{\;0.0181}$ &  $ 0.0491_{\;0.0030}$ \\
        & TTA &  $ 0.9637_{\;0.0011}$ &  $ \mb{0.8598}_{\;0.0105}$ &  $ 0.0558_{\;0.0038}$ \\
        & TTA Ensemble &  $ 0.9654_{\;0.0005}$ &  $ 0.8596_{\;0.0029}$ &  $ 0.0570_{\;0.0014}$ \\
        & SVI &  $ 0.9596_{\;0.0050}$ &  $ 0.8243_{\;0.0115}$ &  $ 0.0586_{\;0.0129}$ \\
        & SVI Ensemble &  $ 0.9635_{\;0.0008}$ &  $ 0.8306_{\;0.0062}$ &  $ 0.0571_{\;0.0020}$ \\
\hline
\multirow{8}{0.14\textwidth}{Uniform}
        & ResNet &  $ 0.9563_{\;0.0037}$ &  $ \mb{0.8383}_{\;0.0215}$ &  $ 0.2678_{\;0.0012}$ \\
        & ResNet Ensemble &  $ 0.9564_{\;0.0039}$ &  $ 0.8325_{\;0.0143}$ &  $ 0.2698_{\;0.0013}$ \\
        & MCDO &  $ 0.9563_{\;0.0034}$ &  $ 0.8074_{\;0.0143}$ &  $ \mb{0.2633}_{\;0.0024}$ \\
        & MCDO Ensemble &  $ \mb{0.9557}_{\;0.0073}$ &  $ 0.7934_{\;0.0228}$ &  $ 0.2644_{\;0.0057}$ \\
        & TTA &  $ 0.9483_{\;0.0047}$ &  $ 0.8294_{\;0.0204}$ &  $ 0.2712_{\;0.0013}$ \\
        & TTA Ensemble &  $ 0.9500_{\;0.0003}$ &  $ 0.8358_{\;0.0092}$ &  $ 0.2719_{\;0.0012}$ \\
        & SVI &  $ 0.9299_{\;0.0072}$ &  $ 0.8010_{\;0.0014}$ &  $ 0.2765_{\;0.0059}$ \\
        & SVI Ensemble &  $ 0.9317_{\;0.0057}$ &  $ 0.8019_{\;0.0030}$ &  $ 0.2770_{\;0.0011}$ \\
\hline
\multirow{8}{0.14\textwidth}{Border}
        & ResNet &  $ 0.9679_{\;0.0016}$ &  $ 0.7879_{\;0.0119}$ &  $ 0.0468_{\;0.0015}$ \\
        & ResNet Ensemble &  $ 0.9739_{\;0.0005}$ &  $ 0.7878_{\;0.0060}$ &  $ 0.0463_{\;0.0015}$ \\
        & MCDO &  $ 0.9706_{\;0.0065}$ &  $ 0.7790_{\;0.0182}$ &  $ 0.0420_{\;0.0025}$ \\
        & MCDO Ensemble &  $ \mb{0.9747}_{\;0.0018}$ &  $ 0.7659_{\;0.0134}$ &  $ \mb{0.0405}_{\;0.0042}$ \\
        & TTA &  $ 0.9719_{\;0.0029}$ &  $ 0.8047_{\;0.0098}$ &  $ 0.0523_{\;0.0053}$ \\
        & TTA Ensemble &  $ 0.9746_{\;0.0024}$ &  $ \mb{0.8107}_{\;0.0038}$ &  $ 0.0500_{\;0.0059}$ \\
        & SVI &  $ 0.9718_{\;0.0058}$ &  $ 0.7922_{\;0.0144}$ &  $ 0.0439_{\;0.0074}$ \\
        & SVI Ensemble &  $ 0.9738_{\;0.0001}$ &  $ 0.7997_{\;0.0002}$ &  $ 0.0532_{\;0.0017}$ \\
\hline
\end{tabular}
\end{table*}

\FloatBarrier
\section{Comparison of uncertainty metrics}
\label{apdx:uncertainty_metrics}

We compare the most commonly used uncertainty metrics of the recent literature, namely confidence, entropy and variance in their ability to distinguish uncertain from certain predictions using the $AUARC$ metric, defined in \autoref{sec:evaluation_settings}. 

\textbf{Confidence:}
A commonly used method, especially in the field of calibration \citep{guo2017}, is to take the maximum of the softmax, which is also called the “confidence” of the prediction. The core idea is that decisions that are considered certain are far away from the decision boundary, while uncertain decisions lie close to the boundary at $1/N$, where $N$ is the number of classes.
\\ \\
\textbf{Entropy:}
Another often used metric \citep{mobiny2019, band2021} is to take the entropy of the prediction as a measure of uncertainty. The entropy of the model's output probability is computed as 
\begin{align}
	H(\mb{y}|\mb{x},\mc{D}) = -\sum_{c \, \in \, \mc{C}} p(y=c|\mb{x}, \mc{D}) \log p(y=c|\mb{x}, \mc{D})
\end{align}
Since the maximum possible entropy varies with the number of classes, we compute the normed entropy as 
\begin{equation}
\begin{aligned}
	H_{\mr{norm}} &= \frac{H}{H_{\mr{max}}} \\
	\text{with} \; H_{\mr{max}} &= - \sum_{i=1}^N \frac{1}{N} \log \left( \frac{1}{N} \right) = \log N
\end{aligned}
\end{equation}
When using the normed entropy, the uncertainty is high when $H_{\mr{norm}} \rightarrow 1$ and uncertainty is low when $H_{\mr{norm}} \rightarrow 0$.

Confidence ($conf$) and entropy behave very similarly in a binary classification setting. Given two predictions $y_i$ and $y_j$ the following relation holds between them:
\begin{equation}
    H(y_i) < H(y_j) \iff conf(y_i) > conf(y_j) 
\end{equation}
\\
\textbf{Variance:}
Since all included uncertainty estimation methods generate a distribution of predictions, the variance of the distribution can be used as a measure of uncertainty \citep{nair2020, leibig2017}. If all predictors agree on a result, the variance is zero, whereas a high variance indicates high uncertainty. 
\\ \\
In \autoref{tab:comp-uncer-metrics} we compare the $AUARC$ values under the strong shift scenario using the three different uncertainty metrics.

As can be seen using the confidence on average produces slightly higher $AUARC$ values, while the relative order of methods stays the same. As the confidence can additionally be used with a singular model, compared to the variance approach we recommend using the confidence as uncertainty metric.

We come to the same conclusion using the weak shift setup but omit it here for brevity. The entries for the confidence metric are the same that can be found in \autoref{tab:AUC-MRC}.

\begin{table*}[tbh]
    
    \centering
    \caption{Mean and standard deviation of the area under the curve of the accuracy-reject curves ($AUARC$) presented in \autoref{fig:reject_vs_acc} for the strong domain shift setup. We compare the confidence / entropy of a single prediction, the confidence / entropy of the mean prediction for ensembling-like approaches and the variance over multiple predictions.}
    \begin{tabular}{l|l|ccc} \label{tab:comp-uncer-metrics}
    Unc. Measure & Unc. Method & ID Centers &        OOD (Center 2) &        OOD (Center 4) \\
    \hline
    Confidence / Entropy & ResNet  &  $ 0.906_{\pm0.017}$ &  $ 0.893_{\pm0.031}$ &  $ 0.954_{\pm0.010}$ \\
    \hline
    \multirow{7}{0.12\textwidth}{Confidence / Entropy}  
    &ResNet Ensemble &  $ 0.918_{\pm0.007}$ &  $ 0.932_{\pm0.010}$ &  $ 0.972_{\pm0.004}$ \\
    &MCDO            &  $ 0.886_{\pm0.030}$ &  $ 0.914_{\pm0.012}$ &  $ 0.971_{\pm0.010}$ \\
    &MCDO Ensemble   &  $ 0.905_{\pm0.008}$ &  $ 0.924_{\pm0.012}$ &  $ 0.971_{\pm0.007}$ \\
    &TTA             &  $ 0.932_{\pm0.005}$ &  $ 0.897_{\pm0.025}$ &  $ 0.970_{\pm0.004}$ \\
    &TTA Ensemble    &  $ \mb{0.933}_{\pm0.002}$ &  $ 0.884_{\pm0.044}$ &  $ 0.963_{\pm0.009}$ \\
    &SVI             &  $ 0.895_{\pm0.027}$ &  $ 0.931_{\pm0.025}$ &  $ 0.972_{\pm0.012}$ \\
    &SVI Ensemble    &  $ 0.896_{\pm0.006}$ &  $ \mb{0.949}_{\pm0.003}$ &  $ \mb{0.981}_{\pm0.002}$ \\

    \hline

    \multirow{7}{0.12\textwidth}{Variance}  
    &ResNet Ensemble &  $ 0.914_{\pm0.007}$ &  $ 0.928_{\pm0.011}$ &  $ 0.972_{\pm0.004}$ \\
    &MCDO            &  $ 0.877_{\pm0.033}$ &  $ 0.911_{\pm0.011}$ &  $ 0.968_{\pm0.012}$ \\
    &MCDO Ensemble   &  $ 0.898_{\pm0.010}$ &  $ 0.915_{\pm0.011}$ &  $ 0.969_{\pm0.008}$ \\
    &TTA             &  $ \mb{0.926}_{\pm0.005}$ &  $ 0.877_{\pm0.031}$ &  $ 0.965_{\pm0.005}$ \\
    &TTA Ensemble    &  $ \mb{0.926}_{\pm0.002}$ &  $ 0.861_{\pm0.053}$ &  $ 0.959_{\pm0.010}$ \\
    &SVI             &  $ 0.890_{\pm0.026}$ &  $ 0.929_{\pm0.023}$ &  $ 0.970_{\pm0.013}$ \\
    &SVI Ensemble    &  $ 0.888_{\pm0.007}$ &  $ \mb{0.942}_{\pm0.004}$ &  $ \mb{0.981}_{\pm0.001}$ \\
    
    \hline
    \end{tabular}

\end{table*}

\FloatBarrier
\section{Ranking of methods across splits}
\label{apdx:camelyon-leave-out}

In addition to defining the weak and strong shift split for the Camelyon17 dataset, we define a leave-one-out scenario, where we train all methods on four centers and evaluate the performance on the remaining out-of-distribution center.
As the strength of the domain shift and therefore the drop in predictive performance between different splits of the centers can differ significantly, it is not possible to directly compare the resulting scores. Instead, we rank all methods for every leave-one-out experiment.

In \autoref{tab:leave-out} you can see the ranking of each method for the metrics $AUARC$, Balanced Accuracy and ECE. 
As can be seen when looking through the columns, no method consistently performs best, instead many methods often claim a high as well as a low rank, further strengthening our results, that there is no best method for uncertainty estimation and robustness on histopathological data.

\begin{table*}[bh]
\centering
\caption{Ranking of the methods for the metrics $AUARC$, Balanced Accuracy and ECE, across all five splits leaving one center out as OOD data. We report the rankings on the ID and OOD data.}
\begin{tabular}{l|l|l|cccccccc} \label{tab:leave-out}
Split            & Metric    & leave &  MCDO &  MCDO &  ResNet &  ResNet  &  SVI &  SVI &  TTA &  TTA \\

& & out & & Ens. & & Ens. & & Ens. && Ens.\\

\hline
\hline

\multirow{15}{*}{ID Centers} &  & 0 &     5 &              2 &       3 &                7 &    8 &             1 &    6 &             4 \\
            &     & 1 &     3 &              7 &       5 &                6 &    2 &             8 &    4 &             1 \\
            & AUARC   & 2 &     5 &              6 &       2 &                3 &    7 &             8 &    1 &             4 \\
            &     & 3 &     5 &              6 &       2 &                3 &    8 &             7 &    1 &             4 \\
            &     & 4 &     5 &              4 &       1 &                6 &    8 &             7 &    2 &             3 \\
\cline{2-11}
            &  & 0 &   5 &            3 &     6 &              1 &  8 &           2 &  7 &           4 \\
            &     & 1 &   2 &            6 &     7 &              8 &  1 &           3 &  5 &           4 \\
            &  Balanced Accuracy   & 2 &   8 &            4 &     5 &              2 &  6 &           7 &  1 &           3 \\
            &     & 3 &   8 &            6 &     7 &              4 &  3 &           2 &  1 &           5 \\
            &     & 4 &   5 &            6 &     3 &              4 &  8 &           7 &  1 &           2 \\
\cline{2-11}
            &  & 0 &     5 &              7 &       1 &                3 &    2 &             8 &    4 &             6 \\
            &     & 1 &     2 &              6 &       1 &                3 &    5 &             7 &    8 &             4 \\
            & ECE    & 2 &     2 &              4 &       5 &                7 &    6 &             8 &    1 &             3 \\
            &     & 3 &     5 &              6 &       1 &                3 &    7 &             8 &    2 &             4 \\
            &     & 4 &     2 &              4 &       1 &                3 &    8 &             7 &    6 &             5 \\
\hline
\multirow{15}{*}{OOD Centers} &  & 0 &     6 &              7 &       2 &                5 &    1 &             8 &    4 &             3 \\
            &     & 1 &     4 &              5 &       3 &                1 &    2 &             6 &    8 &             7 \\
            &   AUARC  & 2 &     8 &              6 &       5 &                2 &    3 &             1 &    7 &             4 \\
            &     & 3 &     8 &              6 &       1 &                4 &    7 &             3 &    2 &             5 \\
            &     & 4 &     8 &              5 &       1 &                4 &    7 &             6 &    2 &             3 \\
\cline{2-11}
            &  & 0 &   8 &            4 &     2 &              6 &  3 &           5 &  7 &           1 \\
            &     & 1 &   3 &            2 &     7 &              5 &  1 &           6 &  8 &           4 \\
            & Balanced Accuracy    & 2 &   8 &            4 &     3 &              1 &  2 &           5 &  6 &           7 \\
            &     & 3 &   7 &            4 &     3 &              5 &  8 &           6 &  1 &           2 \\
            &     & 4 &   1 &            7 &     8 &              6 &  4 &           2 &  5 &           3 \\
\cline{2-11}
                        &  & 0 &     3 &              6 &       1 &                4 &    2 &             7 &    5 &             8 \\
            &     & 1 &     2 &              6 &       3 &                4 &    1 &             5 &    8 &             7 \\
            &  ECE   & 2 &     8 &              4 &       7 &                2 &    5 &             1 &    3 &             6 \\
            &     & 3 &     1 &              4 &       2 &                3 &    7 &             8 &    6 &             5 \\
            &     & 4 &     2 &              4 &       1 &                3 &    6 &             8 &    5 &             7 \\
\hline
\hline
\end{tabular}

\end{table*}

\end{document}